\documentclass{article}
\usepackage[T1]{fontenc}
\usepackage[utf8]{inputenc}
\usepackage{xcolor}
\usepackage{array}
\usepackage{float}
\usepackage{calc}
\usepackage{multirow}
\usepackage{amsmath}
\usepackage{graphicx}
\usepackage{microtype}
\usepackage[unicode=true,
 bookmarks=false,
 breaklinks=false,pdfborder={0 0 1},backref=section,colorlinks=false]
 {hyperref}

\makeatletter

\providecommand{\tabularnewline}{\\}

\usepackage[preprint]{neurips_2023}
\usepackage{soul}
\usepackage{url}% simple URL typesetting
\usepackage{booktabs}% professional-quality tables
\usepackage{amsfonts}% blackboard math symbols
\usepackage{nicefrac}% compact symbols for 1/2, etc.
\usepackage{xcolor}% colors
\usepackage{physics}
\usepackage{subcaption}
\usepackage{algorithm,algpseudocode}
\usepackage{matlab-prettifier}

\title{Benchmarking Automatic Speech Recognition coupled LLM Modules for Medical Diagnostics}

\author{%
  Kabir Kumar\\
  \texttt{ee3210741@iitd.ac.in} \\
	Indian Institute of Technology, Delhi
}

\makeatother

\usepackage{listings}

\begin{document}
\maketitle
\begin{abstract}
Natural Language Processing (NLP) and Voice Recognition agents are
rapidly evolving healthcare by enabling efficient, accessible, and
professional patient support while automating grunt work. This report
serves as my self project wherein models finetuned on medical call
recordings are analysed through a two-stage system: Automatic Speech
Recognition (ASR) for speech transcription and a Large Language Model
(LLM) for context-aware, professional responses. ASR, finetuned on
phone call recordings provides generalised transcription of diverse
patient speech over call, while the LLM matches transcribed text to
medical diagnosis. A novel audio preprocessing strategy, is deployed
to provide invariance to incoming recording/call data, laden with
sufficient augmentation with noise/clipping to make the pipeline robust
to the type of microphone and ambient conditions the patient might
have while calling/recording. Find the deployed pipeline here: \texttt{\textcolor{blue}{\href{https://huggingface.co/spaces/Kabir259/medspeechrec}{https://huggingface.co/spaces/Kabir259/medspeechrec}}}
\end{abstract}

\section{Introduction}

In the critical field of healthcare, accurate medical transcription
is far more than an administrative task---it is fundamental to effective
patient care and informed treatment planning {[}1, 2{]}. Automatic
Speech Recognition (ASR) systems, increasingly adopted in clinical
environments, are designed to convert spoken interactions into precise
written records {[}2, 3{]}. However, these systems face persistent
challenges. Capturing the intricacies of medical conversations, which
often include diverse accents, regional dialects, and specialized
medical terminology, remains a significant hurdle {[}4, 5{]}. The
complexity is heightened in clinical settings, where accurately interpreting
subtle expressions and technical terms is vital to ensuring clarity
and precision in patient records.\\
Errors in medical ASR systems are diverse and problematic, ranging
from misinterpreted drug names and dosages to incorrect lab values,
anatomical confusions, age and gender mismatches, and even wrong doctor
names or dates {[}6{]}. Additional issues include the generation of
nonsensical words, as well as omissions and duplications {[}6, 7{]}.
These inaccuracies can have serious implications, potentially compromising
patient diagnoses and treatment decisions {[}8{]}. Overcoming these
limitations requires innovative solutions beyond the current capabilities
of traditional ASR systems.\\
Large Language Models (LLMs), trained on massive text datasets, these
models exhibit an exceptional ability to understand, contextualize,
and interpret language with high precision {[}9, 10{]}. Recent research
has explored integrating LLMs with audio encoders for direct speech
recognition, expanding their potential applications in ASR and opening
new possibilities for addressing these critical challenges {[}11,
12, 13{]}.\\
Poor-quality call recordings, often affected by noise, clipping, and
compression, significantly degrade ASR performance, leading to inaccurate
transcriptions and reduced reliability {[}14,15{]}. Addressing these
challenges requires effective audio signal processing, particularly
through filtering and equalization techniques {[}16{]}.\\
Low-pass and high-pass filters in digital signal processing (DSP)
have proven effective in isolating critical speech frequencies while
suppressing unwanted noise {[}15,17{]}. Low-pass filtering reduces
high-frequency noise, while high-pass filtering minimizes low-frequency
hums {[}16,17{]}. Equalization techniques further enhance audio intelligibility
by correcting uneven frequency responses, particularly common in compressed
or distorted call recordings{[}14{]}.\\
Research highlights the direct benefits of these techniques for ASR
systems, showing improved transcription accuracy by mitigating noise,
clipping, and other distortions {[}14,15{]}. By enhancing the clarity
of call recordings, these DSP methods optimize audio for ASR processing,
addressing the unique challenges of noisy and compressed environments
{[}16{]}.

\section{Problem Formulation}

\begin{figure}

\begin{centering}
\includegraphics[width=15cm,height=15cm,keepaspectratio]{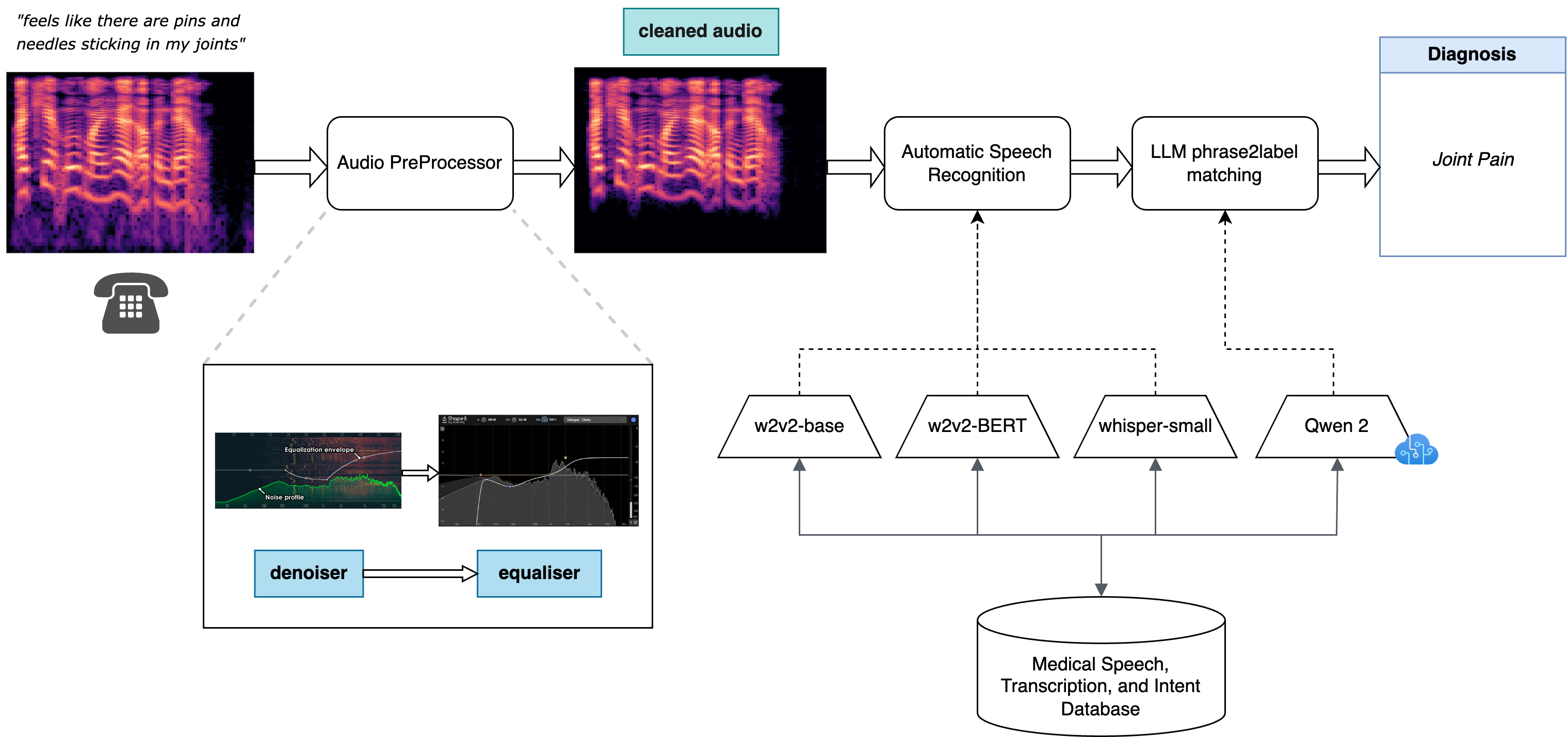}\caption{Proposed Framework}
\par\end{centering}
\end{figure}

Given a noisy audio signal \( S \), the denoising process transforms it into a denoised signal \( S' \):
\[ S' = \text{denoising}(S) \]
The denoised signal \( S' \) is further processed through an equalization step to apply specific filtering operations, resulting in the cleaned audio signal \( S'' \):
\[ S'' = \text{equalized}(S') \]
where the equalization process applies the following filters: \begin{itemize}     \item A high-pass filter centered at 250 Hz,     \item A low-pass filter centered at 11,000 Hz,     \item A high-shelf filter centered at 4,000 Hz. \end{itemize}
Let the cleaned audio signal \( S_0 \) be:
\[ S_0 = S'' \]
The cleaned audio signal \( S_0 \) is input to the ASR system, which produces a sequence of transcribed words \( \{T_i\}_{i=1}^N \), where each \( T_i \) represents a word in the transcribed sentence:
\[ \{T_i\}_{i=1}^N = \text{ASR}(S_0) \]
The transcribed sentence \( T \), composed of \( \{T_i\}_{i=1}^N \), is then passed to a language model (LLM) for classification, resulting in a label \( L \):
\[ L = \text{LLM}(T) \]

\subsection*{Objective}

The objective is to optimize the process such that the label \( L \) accurately reflects the intended classification, minimizing the error introduced by noise in \( S \) and ensuring robust performance of the ASR and LLM systems. Latency should be minimal.

\section{Methodology}

\subsection{Audio PreProcessing}

\subsubsection{Noise}

In acoustic terms, noise is typically associated with high-frequency
disturbances and stereo imbalances. However, call recordings are inherently
converted to mono signals during transmission, as telecommunication
systems prioritize bandwidth efficiency and interpretability. This
conversion reduces stereo artifacts but retains high-frequency disturbances,
which may interfere with Automatic Speech Recognition (ASR) systems.
Additionally, mono signals make noise components easier to process
but leave the signal susceptible to other forms of distortion.

\subsubsection{Clipping}

Low-frequency noise, or \textquotedbl rumble,\textquotedbl{} is a
significant contributor to distortion in audio signals. As rumble
is inherently mono, low frequencies cannot exhibit stereo characteristics
due to phase cancellation issues; any stereo representation of such
frequencies would result in destructive interference. When rumble
occupies a large portion of the signal, especially in recordings made
with cheap consumer grade microphones(as the ones present in our phones),
some even featuring bass-boost technology, it can overwhelm the system's
dynamics. This often leads to distortion, where the rumble’s waveforms
flatten or \textquotedbl clip\textquotedbl{} into square-like shapes
upon hitting the volume ceiling or limiter. Expensive microphones
are designed to be more sensitive to high frequencies while managing
low-frequency handling, but cheap alternatives exacerbate this problem,
leading to degraded audio quality.

\subsubsection{Can Equalization Remove Both Noise and Clipping?}

Equalization offers an effective solution for addressing static noise
and clipping in call recordings. A high-pass filter can block low-frequency
rumble, significantly reducing its interference with the signal. Similarly,
a low-pass filter can eliminate static high-frequency noise, such
as hiss or transmission artifacts, which are common in compressed
call recordings. Music producers often use an additional high-shelf
filter to enhance vocal clarity, boosting high frequencies by a few
decibels to introduce \textquotedbl shimmer\textquotedbl{} and improve
perceived audio quality. This technique is adapted here to clean spoken
audio for further use in the ASR model.

\subsubsection{On AI Denoising Models}

While equalization can address static noise---largely caused by poor
equipment or transmission---dynamic noise presents a more complex
challenge. Dynamic noise, such as crowd chatter or ambient environmental
sounds, shifts across the frequency spectrum, making it harder to
isolate with simple filtering techniques. In these cases, AI-driven
denoising models are much better. These models can adapt to varying
noise profiles, identifying and suppressing unwanted components in
real-time without compromising the primary signal. This capability
makes AI denoising better compared to traditional equalization. However
in our case, it is assumed that noise remains static (only due to
recording and transmission artifacts). Use of AI models, however add
latency to the pipeline which makes it slow. 

\subsection{ASR}

Whisper, trained on a large, supervised multilingual and multitask
dataset, delivers robust out-of-the-box transcription, excelling in
noisy and diverse environments, such as medical dictations. In contrast,
wav2vec 2.0 uses self-supervised learning to extract speech representations
from unlabeled data, requiring fine-tuning to adapt to domain-specific
tasks like medical terminology. Whisper provides immediate usability
without additional training. With minimal finetuning (albeit longer
training times), Whisper outperforms (check results) wav2vec 2.0 as
it is has a transformer based architecture, which is magnitudes better
than a CNN based archtiecture like that of wav2vec 2.0.

\subsection{LLM}

Qwen2 and Llama3 are both famous open source LLMs, however I choose
to use Qwen2 due to:
\begin{table}[H]
\centering{}%
\begin{tabular}{|c|c|c|}
\hline 
Feature & Qwen2 & Llama3\tabularnewline
\hline 
\hline 
Speed & 7-24\% faster than Llama3  & Slower, particularly in complex tasks \tabularnewline
\hline 
Tokens per Second & 11-16 tokens/second  & \textasciitilde 3x slower than Qwen 2 \tabularnewline
\hline 
Context Length  & Up to 128K tokens  & Shorter context length \tabularnewline
\hline 
Multilingual Support  & Strong, responds with the query language & Limited, often defaults to English \tabularnewline
\hline 
\end{tabular}\caption{Comparison between Qwen2 and LLama3}
\end{table}

Other parameters like pricing etc. are similar for both models. The
main differentiating factor for choosing Qwen2 is its better \href{https://aimlapi.com/comparisons/qwen-2-vs-llama-3-comparison}{speed}
and its \href{https://www.leadergpu.com/articles/532-qwen-2-vs-llama-3}{performance in NLP tasks}
as compared to LLama3.

\section{Experiments}

\subsection{Dataset}

The \href{https://www.kaggle.com/datasets/paultimothymooney/medical-speech-transcription-and-intent}{Medical Speech, Transcription, and Intent}
Dataset is used as mentioned.
\begin{center}
\[
\begin{array}{c|ccc}
\hline \text{ Total Samples} & \text{Train} & \text{Test} & \text{Validation}\\
\hline 6661 & 381 & 5895 & 385
\\\hline \end{array}
\]
\par\end{center}

There are multiple fields for each sample detailing the Quietness,
Clipping and Noise in the corresponding audio snippet of the entry.
We crop this information out as the Audio PreProcessor module handles
that on its own. 

\subsubsection{ASR}

For finetuning the \textbf{ASR} module, irrelevant fields were removed
and we were left with a consolidated representation of the dataset
in dictionary format which looked like this:

\inputencoding{latin9}\begin{lstlisting}[language=Matlab,basicstyle={\scriptsize\ttfamily}]
DatasetDict({
    train: Dataset({
        features: ['text', 'audio'],
        num_rows: 381
    })
    test: Dataset({
        features: ['text', 'audio'],
        num_rows: 5895
    })
    validate: Dataset({
        features: ['text', 'audio'],
        num_rows: 385
    })
})
\end{lstlisting}
\inputencoding{utf8}where the text field corresponds to the \textit{phrase} field in the
original csv file and the audio field is extracted from the subdirectory
given.

\subsubsection{LLM}

Similarly, for finetuning the \textbf{LLM} module,the dataset in dictionary
format which looked like this:

\inputencoding{latin9}\begin{lstlisting}[basicstyle={\scriptsize\ttfamily}]
DatasetDict({
    train: Dataset({
        features: ['sentence', 'label'],
        num_rows: 381
    })
    test: Dataset({
        features: ['sentence', 'label'],
        num_rows: 5895
    })
    validate: Dataset({
        features: ['sentence', 'label'],
        num_rows: 385
    })
})
\end{lstlisting}
\inputencoding{utf8}where the sentence field corresponds to the \textit{phrase} field
and the label field corresponds to the \textit{prompt} field in the
original csv file.\\
This was then transformed to the \textbf{\href{https://huggingface.co/datasets/tatsu-lab/alpaca}{Alpaca format}
}which is required for Qwen2/LLama3 to be trained on. It looks like:\inputencoding{latin9}
\begin{lstlisting}[basicstyle={\scriptsize\ttfamily}]
Dataset({
    features: ['instruction', 'input', 'output', 'text'],
    num_rows: 999
})
\end{lstlisting}
\inputencoding{utf8}and every sample looks like:\inputencoding{latin9}
\begin{lstlisting}[basicstyle={\scriptsize\ttfamily}]
alpaca_prompt = """Given a sentence generated via a Speech to Text model, 
clean the sentence grammatically and make it sound natural. Then classify 
the speaker's medical conditon in the given sentence.

### Instruction:
{}

### Input:
{}

### Response:
{}
<|endoftext|>"""
\end{lstlisting}
\inputencoding{utf8}

\subsection{Parameter Tuning and Libraries}

\subsubsection{ASR}

For ASR, a learning rate of 5e-5 , weight decay of 0.005, and constant,
linear and cosine learning rate schedule were experimented with. The
models usually got finetuned in 1500-2000 steps. Warmup steps, set
to 200 (approximately 10\% of the 1,000 total training steps).\\
Training was conducted using mixed-precision (fp16) to address limited
computational resources. Mixed-precision training significantly reduces
memory consumption and computational load, enabling faster processing
and larger batch sizes. Most operations in fp16 are sufficient for
convergence while critical steps are retained in fp32.\\
The word error rate (WER) was selected from the \texttt{jiwer} library
as the primary evaluation metric due to its direct relevance to ASR
tasks. Unlike general-purpose metrics such as accuracy or F1 score,
WER evaluates transcription quality by measuring the total number
of insertions, deletions, and substitutions required to match predicted
transcriptions with ground truth, keeping in mind minor transcription
inaccuracies can significantly impact usability.

\subsubsection{LLM}

For the LLM, I used the \texttt{unsloth }library to significantly
speed up the training process. I ran it for 20 epochs with 5 warmup
steps with a linear learning rate sceduler and used LoRA PEFT method.
Since I had more leeway with compute here (due to the training library
I used), I opted for bf16 training instead of fp16 training (I didn't
increase the epochs/steps or learning rate as the model started oscillating
in loss later on). bf16 is better than fp16 as it has a wider dynamic
range due to its 8-bit exponent (compared to fp16’s 5-bit), allowing
it to handle larger and smaller values more effectively without overflow
or underflow. 

\subsection{Results}

\subsubsection{ASR}

\begin{table}[H]

\begin{centering}
\begin{tabular}{|c|c|c|c|c|}
\hline 
\multirow{2}{*}{Model} & \multirow{2}{*}{Version} & \multicolumn{2}{c|}{WER(\%)} & \multirow{2}{*}{Steps Taken To Train}\tabularnewline
\cline{3-4} \cline{4-4} 
 &  & Validation & End of Training & \tabularnewline
\hline 
\hline 
\multirow{4}{*}{wav2vec2} & base & 135 & - & -\tabularnewline
\cline{2-5} \cline{3-5} \cline{4-5} \cline{5-5} 
 & base-\textbf{FT} & 48.9 & 32.8 & 2500 (2.5 hrs)\tabularnewline
\cline{2-5} \cline{3-5} \cline{4-5} \cline{5-5} 
 & BERT & 100 & - & -\tabularnewline
\cline{2-5} \cline{3-5} \cline{4-5} \cline{5-5} 
 & BERT-\textbf{FT} & 37.5 & 23.1 & \textbf{500} (1 hr)\tabularnewline
\hline 
\multirow{2}{*}{Whisper} & small(3B) & 128 & - & -\tabularnewline
\cline{2-5} \cline{3-5} \cline{4-5} \cline{5-5} 
 & small(3B)-\textbf{FT} & \textbf{\textcolor{red}{21.3}} & 9.97 & 1000 (2.5 hrs)\tabularnewline
\hline 
\end{tabular}\caption{Results for ASR(Whisper and wav2vec 2.0)}
\par\end{centering}
\end{table}

The Whisper model is extremely compute intensive(even the small 3B
version) however, gives the best results as shown. A performance of
21.3\% WER with minimal fitnetuning in Whisper indicated that with
dedicated finetuning with more data and compute, it can quickly scale
up to be an industry ready model with high reliability. Whisper, can
quickly learn new accents and dialects and can learn to differentiate
speech in noisy input, making its finetuning all the more worthwhile.
This goes to shows how Whisper is the best ASR model for our use case,
owing to its huge training library and transformer architecture. \\
It is interesting to note that without finetuning, no model, either
ASR or LLM is able to perform on the dataset. wav2vec2-BERT shows
a significant improvement over wav2vec2-base both in terms of WER
and time taken to finetune, however, can't match Whisper. %
\noindent{\fboxrule 0.8pt\fcolorbox{blue}{white}{\begin{minipage}[t]{1\columnwidth - 2\fboxsep - 2\fboxrule}%
\textbf{Finetuning Whisper is effective and excellent in handling
domain-specific ASR tasks such as Medical Speech Recognition}, given
enough compute, time and data.%
\end{minipage}}}

\subsubsection{LLM}

\begin{table}[H]
\begin{centering}
\begin{tabular}{|c|c|c|c|c|}
\hline 
\multirow{2}{*}{Model} & \multirow{2}{*}{Version} & \multicolumn{2}{c|}{Accuracy(\%)} & \multirow{2}{*}{Steps Taken To Train}\tabularnewline
\cline{3-4} \cline{4-4} 
 &  & Validation & End of Training & \tabularnewline
\hline 
\hline 
\multirow{2}{*}{Qwen2} & 7B & - & - & -\tabularnewline
\cline{2-5} \cline{3-5} \cline{4-5} \cline{5-5} 
 & 7B-\textbf{FT(LoRA)} & 20.0 & 25.5 & 20 (30 mins)\tabularnewline
\hline 
\end{tabular}\caption{Results for LLM(Qwen2)}
\par\end{centering}
\textit{\textcolor{gray}{The LLM model needs to get fine-tuned in
order to generate one-word label classifications. Otherwise it will
just start responding like a general LLM.}}
\end{table}

The training was extremely fast due to LoRA. The reason I opted for
such low steps was due to the oscillating loss of the model when I
ran it for 200 steps earlier. The low validation accuracy shouldn't
be considered '\textbf{that}' dreadful as many sentences wrongly classified
logically seem to fit those labels to some extent too! (\textit{the
issue lies with lack of data, as LLMs are data hungry models}).
\begin{table}[H]
\begin{centering}
\begin{tabular}{|c|c|}
\hline 
\multirow{2}{*}{Predicted} & \multirow{2}{*}{Ground Truth}\tabularnewline
 & \tabularnewline
\hline 
\hline 
stomach ache  & emotional pain \tabularnewline
\hline 
hard to breath  & feeling dizzy\tabularnewline
\hline 
knee pain & injury from sports\tabularnewline
\hline 
stomach ache & feeling dizzy\tabularnewline
\hline 
\end{tabular}\caption{Excerpt of the inference results for the LLM classifier}
\par\end{centering}
\textit{\textcolor{gray}{Check ./Benchmarking/QWEN2\_inf+val.ipynb
for inference and validation results.}}
\end{table}

\section{Conclusion}

This report demonstrates the integration of ASR systems, such as Whisper
and wav2vec 2.0, with LLMs like Qwen2 for medical speech recognition
and diagnostics. Whisper's transformer-based architecture and extensive
training dataset enable it to address challenges posed by noisy environments
and the linguistic complexities of medical contexts effectively.\\
It is found that Qwen2 is suitable for contextualizing transcribed
speech in classification tasks, supported by its processing speed,
extended context length, and multilingual capabilities. While currently
applied to label classification, the scalability of such LLMs suggests
potential for further development into conversational agents or chatbots
capable of assisting in medical consultations and streamlining healthcare
workflows.

\section{References}

{\small{}{[}1{]} J. Zhang, J. Wu, Y. Qiu, A. Song, W. Li, X. Li, and
Y. Liu. Intelligent speech technologies for transcription, disease
diagnosis, and medical equipment interactive control in smart hospitals:
A review. Comput Biol Med, 153:106517, Feb 2023. Epub 2023 Jan 5.}\\
{\small{}{[}2{]} M. Johnson, S. Lapkin, V. Long, P. Sanchez, H. Suominen,
J. Basilakis, and L. Dawson. A systematic review of speech recognition
technology in health care. BMC Med Inform Decis Mak, 14(94):1--18,
Oct 2014. }\\
{\small{}{[}3{]} K. Saxena, R. Diamond, R. F. Conant, T. H. Mitchell,
I. G. Gallopyn, and K. E. Yakimow. Provider adoption of speech recognition
and its impact on satisfaction, documentation quality, efficiency,
and cost in an inpatient ehr. AMIA Jt Summits Transl Sci Proc, pages
186--195, 2017. Epub 2018 May 18.}\\
{\small{}{[}4{]} Li Zhou, Suzanne V. Blackley, Leigh Kowalski, Raymond
Doan, Warren W. Acker, Adam B. Landman, Evgeni Kontrient, David Mack,
Marie Meteer, David W. Bates, and Foster R. Goss. Analysis of Errors
in Dictated Clinical Documents Assisted by Speech Recognition Software
and Professional Transcriptionists. JAMA Network Open, 1(3):e180530--e180530,
07 2018.}\\
{\small{}{[}5{]} Alex DiChristofano, Henry Shuster, Shefali Chandra,
and Neal Patwari. Global performance disparities between english-language
accents in automatic speech recognition. ArXiv, abs/2208.01157, 2023.}\\
{\small{}{[}6{]} Tobias Hodgson and Enrico Coiera. Risks and benefits
of speech recognition for clinical documentation: a systematic review.
Journal of the American Medical Informatics Association, 23(e1):e169--e179,
11 2015.}\\
{\small{}{[}7{]} McGurk S, Brauer K, Macfarlane TV, and Duncan KA.
The effect of voice recognition software on comparative error rates
in radiology reports. Br J Radiol, 81(970):767-- 770, 2008. Epub
2008 Jul 15.}\\
{\small{}{[}8{]} Adane K, Gizachew M, and Kendie S. The role of medical
data in efficient patient care delivery: a review. Risk Manag Healthc
Policy, 12:67--73, Apr 2019.}\\
{\small{}{[}9{]} Humza Naveed, Asad Ullah Khan, Shi Qiu, Muhammad
Saqib, Saeed Anwar, Muhammad Usman, Naveed Akhtar, Nick Barnes, and
Ajmal Mian. A comprehensive overview of large language models. ArXiv,
abs/2307.06435, 2023.}\\
{\small{}{[}10{]} Takeshi Kojima, Shixiang Shane Gu, Machel Reid,
Yutaka Matsuo, and Yusuke Iwasawa. Large language models are zero-shot
reasoners. ArXiv, abs/2205.11916, 2023.}\\
{\small{}{[}11{]} Yassir Fathullah, Chunyang Wu, Egor Lakomkin, Junteng
Jia, Yuan Shangguan, Ke Li, Jinxi Guo, Wenhan Xiong, Jay Mahadeokar,
Ozlem Kalinli, Christian Fuegen, and Mike Seltzer. Prompting large
language models with speech recognition abilities. ArXiv, abs/2307.11795,
2023.}\\
{\small{}{[}12{]} Yukiya Hono, Koh Mitsuda, Tianyu Zhao, Kentaro Mitsui,
Toshiaki Wakatsuki, and Kei Sawada. An integration of pre-trained
speech and language models for end-to-end speech recognition. ArXiv,
abs/2312.03668, 2023.}\\
{\small{}{[}13{]} Paul K. Rubenstein, Chulayuth Asawaroengchai, Duc
Dung Nguyen, Ankur Bapna, Zal\'{ }an Borsos, et al. Audiopalm: A large
language model that can speak and listen. ArXiv, abs/2306.12925, 2023.}\\
{\small{}{[}14{]} S. K. S. Reddy et al., \textquotedbl Impact of
Noise on Automatic Speech Recognition Systems,\textquotedbl{} Journal
of Speech Communication, vol. 98, pp. 1-12, 2021.}\\
{\small{}{[}15{]} H. Wang and D. Wang, \textquotedbl Audio Signal
Processing for Speech Enhancement,\textquotedbl{} IEEE Transactions
on Audio, Speech, and Language Processing, vol. 28, no. 5, pp. 1234-1247,
2020.}\\
{\small{}{[}16{]} Y. Xu et al., \textquotedbl Enhancing ASR Performance
in Noisy Environments Using Signal Processing Techniques,\textquotedbl{}
Computer Speech \& Language, vol. 54, pp. 1-15, 2019. }\\
{\small{}{[}17{]} Equalizers, IIR and FIR https://thesofproject.github.io/latest/algos/eq/equalizers\_tuning.html}{\small\par}
\end{document}